\documentclass[twocolumn,showpacs,preprintnumbers,amsmath,amssymb]{revtex4}

\usepackage{pslatex,graphicx}

\begin{document}

\title{Stability Analysis of a Hybrid Cellular Automaton Model of Cell Colony Growth}

\author{P. Gerlee}
\email{gerlee@maths.dundee.ac.uk}
\author{A.R.A Anderson}
\affiliation{
Division of Mathematics, University of Dundee,\\
DD1 4HN, Scotland
} 

\date{\today}

\begin{abstract}
Cell colonies of bacteria, tumour cells and fungi, under nutrient limited growth conditions,  exhibit complex branched growth patterns. In order to investigate this phenomenon we present a simple hybrid cellular automaton model of cell colony growth. In the model the growth of the colony is limited by a nutrient that is consumed by the cells and which inhibits cell division if it falls below a certain threshold.  Using this model we have investigated how the nutrient consumption rate of the cells affects the growth dynamics of the colony. We found that for low consumption rates the colony takes on a Eden-like morphology, while for higher consumption rates the morphology of the colony is branched with a fractal geometry. These findings are in agreement with previous results, but the simplicity of the model presented here allows for a linear stability analysis of the system. By observing that the local growth of the colony is proportional to the flux of the nutrient we derive an approximate dispersion relation for the growth of the colony interface. This dispersion relation shows that the stability of the growth depends on how far the nutrient penetrates into the colony. For low nutrient consumption rates the penetration distance is large, which stabilises the growth, while for high consumption rates the penetration distance is small, which leads to unstable branched growth. When the penetration distance vanishes the dispersion relation is reduced to the one describing Laplacian growth without ultra-violet regularisation. The dispersion relation was verified by measuring how the average branch width depends on the consumption rate of the cells and shows good agreement between theory and simulations.

\end{abstract}

\maketitle

\section{Introduction}
Pattern formation in living systems has attracted much attention since the pioneering work of D'Arcy Thompson \cite{darcy}. In recent years special attention has been given to patterns emerging from cell colony growth in hostile environments \cite{benreview, benreview2}. These systems tend to exhibit complex growth patterns when the growth is limited by the diffusion of a nutrient that is necessary for the growth of the cells. The morphologies obtained from these living systems resemble that of many non-living systems like electrodeposition \cite{electro}, crystal growth \cite{crystal} and viscous fingers \cite{viscous, porous}. All of these non-living systems obey the same underlying growth principle, which is that of Laplacian growth, in which the interface between the two phases is advanced at a rate proportional to the gradient of a potential field. In the case of electrodeposition it is the electric field around the substrate, in crystal growth it is the temperature field and in viscous fingering the pressure in the liquid. This similarity between biological and non-living diffusion limited patterns has led to the hypothesis that the biological patterns could be explained with the same basic principles \cite{mats}.

Perhaps the most studied example of cell colony growth is the growth of bacterial colonies subject to low nutrient levels. Bacteria are usually grown in petri dishes at high nutrient concentration. These conditions give rise to colonies with simple compact morphologies, but when the growth occurs in more hostile low nutrient concentrations the morphologies of the colonies take on very complex shapes. This phenomena was first reported by Matsushita et al. \cite{mats} and since then several models have been suggested to explain the observed mophologies. The main modelling approach that has been used is to consider the growth via a system of reaction-diffusion equations \cite{mats2, mimura, kitsu, kawa}. These models are able to reproduce the observed patterns, ranging from Eden-like \cite{Eden} and dense branched morphologies \cite{DBM} to DLA-like patterns \cite{DLA}. Another approach by Ben-Jacob et al. \cite{walker, walker2} is to model the bacteria as clusters of discrete walkers which obey dynamical rules. This model also agrees well with experimental data and is more biologically realistic compared to the reaction-diffusion approach.

Avascular tumours also grow under similar nutrient limited conditions as bacteria cultured in petri dishes. In the early stages of cancer development the tumour has yet to acquire its own vasculature and the cancer cells therefore have to rely on diffusion as the only means of nutrient transport \cite{suth}. When the tumour reaches a critical size the diffusion of nutrients is not enough to supply the inner parts of the tumour with oxygen, this leads to cell death or necrosis in the core of the tumour. Surrounding the necrotic core is a rim of quiescent cells and on the outer boundary there is a thin rim of proliferating cells. The mitotic activity therefore only takes place in a small fraction of the tumour, while the majority of the tumour consists of cells that are either quiescent or dead. Although the growth of a tumour is a much more complex process compared to the growth of bacteria in petri dishes there is still evidence from both experiments \cite{vilela, cristini, cristini2} and mathematical models \cite{gerlee, ferreira, anderson, cell} that tumours exhibit fingering and fractal morphologies driven by diffusion limited growth.

Another biological system that displays complex patterns under diffusion limited growth are fungal colonies. Complex patterns with fractal morphologies have been observed for both multi-cellular filamentous growth \cite{matsuura} and for yeast-like unicellular growth \cite{sams}. These patterns arise in low nutrient conditions or when there is a build up of metabolites which inhibit the growth of the colony and have successfully been modelled using both continuous \cite{dyce, regalado} and discrete \cite{lopez, sams} techniques.

Bacterial colonies exhibit branches which have a width of approximately 0.5 mm, which is of the order of 100 cells \cite{mats2}. This is very different from viscous fingers for example, where the disparity of length scales between the molecules and pattern is much larger. We believe that in order to fully capture the dynamics of such systems, where the characteristic length scale of the pattern is similar to that of the cells which constitute the pattern, it is necessary to model them at the level of single cells. In this paper we therefore present a simple hybrid cellular automaton model of nutrient dependent cell colony growth where each automaton element represents a single cell. The aim of this model is not to represent any specific biological system, but rather to show that complex growth patterns can emerge from a very simple model with minimal assumptions about the cell behaviour. The simplicity of the model presented in this paper ensures both generality of the results discussed as well as allowing us to carry out a stability analysis. This analysis we hope will shed light on the growth instabilities observed in the above mentioned systems.

\section{The Model}

The domain of the colony is restricted to a 2-dimensional substrate and the growth is assumed to be limited by some essential nutrient which is required for cell division. The substrate on which the cells grow is represented by a $N\times N$ cellular automaton with lattice constant $\Delta$. Each automaton element can be in three different states: (i) empty, (ii) hold an active cell or (iii) hold an inactive cell and each element is identified by a coordinate $\vec x=\Delta(i,j)$ $i,j=0,1,2,...,N-1$. The cellular automaton is coupled with a continuous field $c(\vec x,t)$ that describes the nutrient concentration. In the case of bacteria the nutrient represents peptone, for tumours oxygen and for fungi some sort of carbon source like glucose. The transition from an active cell to an inactive occurs if the nutrient concentration falls below some critical threshold $c_n$. This inactive state corresponds to sporulation or cell quiescence and is assumed to be irreversible. An active cell divides when it has reached maturation age, it then places a daughter cell at random in an empty neighbouring grid point (using a von Neumann neighbourhood). If none of the neighbouring grid points are empty the cell division fails, but the cell remains in the active state. After cell division has occurred the age of the parent cell is reset, which means that is has to reach maturation age again to divide. In order to account for variation in maturation age between different cells the maturation age of each cell is chosen randomly from a $N(\tau,\sigma)$ normal distribution, where $\tau$ represents the average maturation age and the variance is set to $\sigma=\tau/2$. For simplicity we will consider non-motile cells (which for bacteria corresponds to high agar concentrations \cite{mats}), which implies that the growth of the colony is driven by cell division.
 
Active cells are assumed to consume nutrients at some fixed rate $k$, while inactive cells don't consume any nutrients. The nutrient is assumed to diffuse in the substrate with a diffusion constant $D$. The nutrient concentration field therefore obeys the equation,
\begin{equation}\label{eq:diffeq}
\frac{{\partial c(\vec x,t)}}{{\partial t}} = D\nabla ^2 c(\vec x,t) - k n(\vec x,t)
\end{equation}
where $n(\vec x,t)=1$ if the automaton element at $\vec x$ holds an active cell and $n(\vec x,t)=0$ if it holds an inactive cell or is empty. The boundary conditions satisfied by the nutrient fields are Dirichlet with a constant value $c_\infty$. This represents a continuous and fixed supply of nutrient from the boundary of the system.
In order to simplify the system we introduce new non-dimensional variables given by,
\begin{equation}\label{eq:nondim}
\vec x'=\frac{\vec x}{\Delta}, \ \ t'=\frac{Dt}{\Delta^2}, \ \ c'=\frac{c}{c_\infty}, \ \ k'=\frac{k\Delta^2}{c_\infty D}.
\end{equation}
Using these new variables the equation describing the nutrient concentration becomes (omitting the primes for notational convenience)

\begin{equation}\label{eq:diffeqnon}
\frac{{\partial c(\vec x,t)}}{{\partial t}} = \nabla ^2 c(\vec x,t) - k n(\vec x,t).
\end{equation}
This equation is discretised using standard five-point finite central difference formulas  and solved on a grid with the same spatial step size as the cellular automaton. Each time step of the simulation the nutrient concentration is solved using the discretised equation and all the active cells on the grid are updated in a random order.

\subsection{Simulations}
\begin{figure}[!hbp]
   \centerline{\includegraphics[width=7cm]{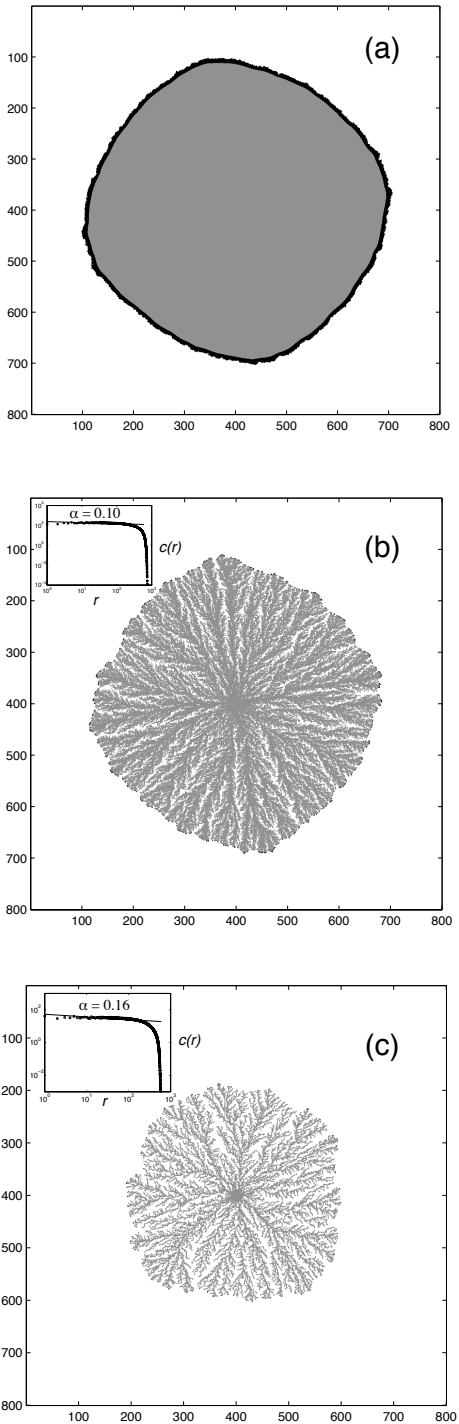}}
\caption{\label{fig:col}Cell colony plots after 300 cell generations for nutrient consumption rates: (a) $k=0.01$, (b) $k=0.1$ and (c) $k=0.2$. The other parameters were fixed at $\tau=10$, $c_n=0.1$ and grid size $N=800$. Empty CA elements are coloured white, inactive cells grey and active cells black. This shows that the colony morphology depends on the nutrient consumption rate, where a high consumption rate gives rise to a fractal colony morphology. The insets in (b) and (c) shows a loglog-plot of the density-density correlation function, which show that at small length scales $C(r) \sim r^{-\alpha}$ in the fractal growth regime.}
\end{figure}

Using this simple model of cell colony growth we have investigated how the nutrient consumption rate $k$ of the cells affects the growth dynamics of the colony. Note that varying the non-dimensional consumption rate $k$ is equivalent to either varying the dimensional consumption rate or the boundary concentration $c_\infty$, see eq. (\ref{eq:nondim}).
All simulations were started with an initial circular colony (with radius 10 cells) of active cells at the centre of the grid and an initial homogeneous nutrient concentration of $c(\vec x,t)=1$. Figure \ref{fig:col} shows the colony after 300 cell generations for $k=0.01, 0.1$ and $0.2$ with $\tau=10$, $c_n=0.1$ on a grid of size $N=800$. 
From this figure it is obvious that the consumption rate of the cells affects the morphology of the colony. For the lowest consumption rate $k=0.01$ the colony grows with a compact Eden-like morphology. The colony consists mostly of inactive cells with an outer rim of active cells at the boundary.
For $k=0.1$ the morphology is no longer compact but exhibits a pattern similar to the dense branching morphology (DBM) observed in viscous fingering \cite{DBM}. Again the colony consists mostly of inactive cells and the few active cells reside on the tips of the branches. For the highest consumption rate $k=0.2$ the branched morphology is even more evident and it exhibits thinner branches. 
In order to characterise the morphologies further we measured the fractal dimension of the colonies by measuring how the number of cells $N$ (active and inactive) depends on the radius $R$ of the pattern \cite{mandel}. For a compact morphology we expect that $N \sim R^2$, which is what we find for $k=0.01$, but for the two other morphologies we find that $N\sim R^{D}$, where $D \approx 1.91$ for $k=0.1$ and $D \approx1.83$ for $k=0.2$. For both $k=0.1$ and $0.2$ the colony thus grows with a fractal morphology. This was also confirmed by measuring the density-density correlation function $C(r)=\langle \rho (r') \rho (r+r')\rangle$ for the colonies (see inset of fig. \ref{fig:col}.b and c). At small length scales $C(r)$ decays as $r^{-\alpha}$, where the fractal dimension of the colony is given by $D=2-\alpha$. For $k=0.1$ we find $\alpha = 0.10$ and for $k=0.2$ we have $\alpha=0.16$, which gives fractal dimensions in close agreement with the previous method.

The Eden-like growth pattern observed for $k=0.01$ is to be expected, as all cells on average divide at uniform speed, but what is interesting is that as the nutrient consumption rate is increased it leads to a branched morphology. The intuitive explanation of why this type of growth occurs is because the high nutrient consumption can't sustain the growth of a smooth colony boundary. If a cell on the boundary divides and places the daughter cell outside the existing colony boundary it reduces the chances of neighbouring cells to divide, as the daughter cell "steals" nutrient at the expense of the cells that are closer to the centre of the colony, effectively creating a screen from access to the nutrients. It is this screening effect that amplifies perturbations to the colony interface and leads to the branched morphology. This implies that the branched colony morphology is a result of nutrient limited growth, which is in agreement with the previously discussed experiments and models of colonies of bacteria, tumour cells and fungi.

The dynamics of this model are essentially that of a diffusion limited growth process, in which the diffusing nutrient is transformed by the cells into biomass in the form of daughter cells. It is therefore not surprising that it exhibits morphologies similar to non-living diffusion limited growth phenomena like viscous fingers, electrodeposition and crystal growth. In the next section we will quantify the growth instabilities observed in the system by performing a linear stability analysis on the model.
\section{Stability Analysis}
In order to analyse the stability of the discrete model we have to construct an analogous model that captures the essential dynamics but is amenable to mathematical analysis. 
\subsection{Sharp interface model}
\begin{figure}[!hbp]
   \centerline{\includegraphics[width=8cm]{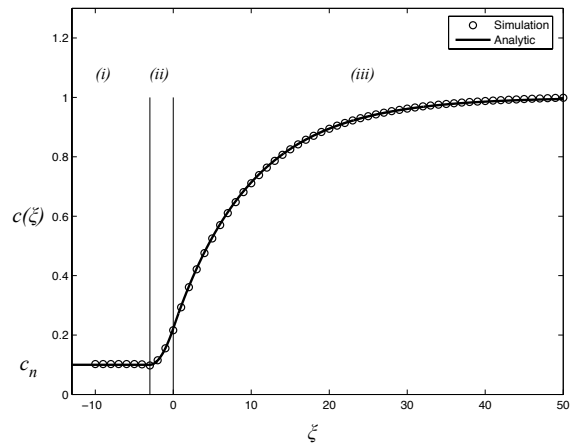}}
\caption{\label{fig:conc}The nutrient profile plotted in the moving $\xi$-frame for $k=0.03$, $c_n=0.1$ and $\tau=10$ in $(i)$ the inactive part $(ii)$ the active part and $(iii)$ outside the colony. The circles represent the radial nutrient profile from a simulation and the solid line is the analytic nutrient profile (\ref{eq:sol}). 
}
\end{figure}

The analogous model will be constructed by considering the colony boundary as a sharp interface that moves in two dimensions with a velocity $v_p$ determined by the maturation age $\tau$ and the size of the individual cells $\Delta=1$, such that $v_p=1/ \tau$. The nutrient consumption of the cells is taken to be $k$ in the active part of the colony, where $c>c_n$, zero in the inactive part, where $c \leq c_n$, and zero outside the colony. If we consider the growth of a planar front, growing in the $y$-direction and stretching infinitely in the $x$-direction, the equations describing the nutrient concentration are given by,
\begin{equation}\label{eq:init1}
\frac{{\partial c(\vec x,t)}}{{\partial t}} = \nabla ^2 c(\vec x,t), \ \ \ y>y_i
\end{equation}
\begin{equation}\label{eq:init2}
\frac{{\partial c(\vec x,t)}}{{\partial t}} = \nabla ^2 c(\vec x,t) - kH(c(\vec x,t)-c_n), \ \ \ y \leq y_i
\end{equation}
where $H$ is the Heaviside step-function and $y_i$ is the position of the interface along the $y$-axis. In order to make the analysis simpler we make a change of coordinates to a moving frame that travels along with the interface, i.e. we define a new coordinate $\xi=y-v_pt$, where $\xi=0$ corresponds to $y=y_i$, the position of the interface. This change of coordinates plus the fact that there is no dependence on $x$ reduces (\ref{eq:init1} -\ref{eq:init2}) to a system of three ODE's given by,
\begin{equation}\label{eq:ode1}
c''+v_pc'=0, \ \ \xi > 0
\end{equation}
\begin{equation}\label{eq:ode2}
c''+v_pc'-k=0, \ \ \ -d < \xi \leq 0
\end{equation}
\begin{equation}\label{eq:ode3}
c''+v_pc'=0, \ \ \ \xi \leq -d 
\end{equation}
where (\ref{eq:ode1}) describes the nutrient concentration outside the colony (\ref{eq:ode2}) in the active region of the colony and (\ref{eq:ode3}) in the inactive region, and where the width of the active region $d$ is determined from the solution of the ODE's. The boundary condition for the nutrient concentration at $\xi=\infty$ is that it should reach the boundary value $c_\infty=1$. Moreover we want the solution to be smooth across the interface so we require that the solutions to (\ref{eq:ode1}) and (\ref{eq:ode2}) have the same value as do their derivatives at $\xi=0$. We also require that the solutions to (\ref{eq:ode2}) and (\ref{eq:ode3}) take the value $c_n$ at $\xi=-d$ and that the derivative is zero at that point. If we let $c_e(\xi)$ be the external solution, $c_a(\xi)$ the solution in the active region and $c_i(\xi)$ in the inactive region we formally require that,
\begin{equation}\label{eq:BC}
\begin{array}{l}
c_e (\xi ) \to 1,\ \ \ \rm{as}\ \ \xi  \to \infty  \\ 
c_e (0) = c_a(0) \\ 
c_e '(0) = c_a '(0) \\ 
c_a (-d) = c_i(-d) = c_n \\
c_a'(-d) = c_i'(-d) = 0. \\
 \end{array}
\end{equation}
A solution to (\ref{eq:ode1} - \ref{eq:ode3}) with boundary conditions (\ref{eq:BC}) is given by
\begin{equation}\label{eq:sol}
c(\xi ) = \left\{ \begin{array}{l}
1-\frac{k}{v_p^2}\left(1-e^{-\frac{v_p^2(1-c_n)}{k}}\right) e^{-v_p\xi},\ \ \xi > 0 \\ 
\frac{k}{v_p^2}\left( v_p\xi+e^{-\frac{v_p^2(1-c_n)}{k}} e^{-v_p\xi} + v_p^2/k -1 \right),\ \ -d < \xi \leq 0 \\
c_n,  \ \ \xi \leq -d
 \end{array} \right.
\end{equation}
where $d = (1-c_n)v_p/k$ is the thickness of the boundary layer. An example of the solution with appropriate parameter values can be seen in fig. \ref{fig:conc}, where the circles represents the nutrient profile measured radially in a simulation with corresponding parameter values. The agreement between the two nutrient profiles shows that the planar front approach approximates the nutrient profile very well. 

\subsection{Interface velocity}
We will shortly analyse the stability of this simple model, but before doing so we will discuss the growth dynamics of the cells in more detail. In the hybrid model the cells divide at a uniform speed, only varied by the stochastic difference in maturation age $\tau$. Although this is the case the model still gives rise to interesting growth patterns. The reason behind this is that the cells in the model become inactive if the nutrient concentration falls below the threshold $c_n$. If a cell on the boundary becomes inactive before cell division occurs the interface velocity becomes zero at that point, and the inactive cell may become the starting point of the development of a fjord. This scenario is only possible if the nutrient consumption rate is sufficiently high compared to the nutrient concentration at the boundary. If this is the case the cells on the boundary have to rely on the flux of nutrient in order to survive long enough to go through cell division. Our interpretation of this is that in the limit of high consumption rates the velocity of the interface becomes proportional to the flux of nutrient, a mechanism already suggested by Matsushita \& Fujikawa \cite{mats}. Mathematically this means that the local interface velocity is given by $v(x) \propto \vec n \cdot \nabla c$, where $\vec n$ is the normal of the interface.
This observation will be the basis for our stability analysis, which means that our treatment of the system will not be rigourously related to our model, but rather aimed more at understanding the dynamics of the model from a qualitative point of view.

\subsection{Instability of the interface}
\begin{figure}[!hbp]
   \centerline{\includegraphics[width=5cm]{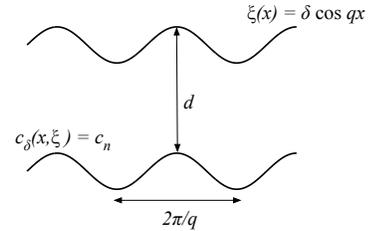}}
\caption{\label{fig:layer}This figure shows the structure of the interface. It is assumed that the curve $c_\delta(x,\xi)=c_n$ is given by displacing the inteface by $-d$ in the $\xi$-direction.}
\end{figure}

In the above solution (\ref{eq:sol}) of (\ref{eq:ode1} - \ref{eq:ode3}) we assumed that the interface at $\xi=0$ was flat, we now introduce a small oscillating perturbation of amplitude $\delta$ to the interface, giving $\xi_0(x,t) = \delta(t) \cos qx$, where $\delta \ll 1$. This changes the nutrient field in the vicinity of the interface, and we need to find this perturbed field $c_\delta(x,\xi)$ to determine the stability. 
In the following analysis we will assume that the interface velocity $v_p \ll 1$, which means that there is a separation in the time-scales between the movement of the interface and the dynamics of the nutrient field. This allows for a number of simplifications: Firstly it implies that the nutrient field is in a quasi-stationary state, which means that the nutrient concentration approximately satisfies $\nabla^2 c = 0$ outside the colony and implies that we can approximate the nutrient profile by a linear function in the vicinity of the interface. This is generally the case for the types of biological system discussed here, where the nutrient diffuses on a time-scale much faster than the growth of the colony. For example the reproduction time of bacterial cell is of the order of hours, while the diffusion constant of nutrient in agar is of the order 10$^{-7}$ cm$^2$/s \cite{golding}. This means that the diffusion time across one bacteria is $\Delta t \approx 0.1$ s (assuming that a bacteria is approx. 10 $\mu$m), which is considerably smaller than the reproduction time. Cancer cells are approximately 25 $\mu$m in diameter and the diffusion constant of oxygen in tissue is $1.8 \times 10^{-5}$ cm$^2$/s \cite{oxy}, giving a diffusion time of $\Delta t \approx 4 \times 10^{-3}$ s across one cell, which is several orders of magnitude smaller than the reproduction time of a cancer cell, which is of the order of 10-20 hours.\\
Secondly the quasi-stationary assumption allows us to omit any time dependence in the solutions for the perturbed field. It also implies that the iso-concentration curve $-d(x)$ defined by $c(x,\xi)=c_n$ will be stationary in the moving frame. Further we will assume that this curve is given by displacing the interface by $-d$ in the $\xi$-direction, i.e $d(x)=d-\delta \cos qx$ (cf. fig. \ref{fig:layer}). This is of course only valid when $d$ is small and when the wave number $q$ of the oscillation is small. The values of $d$ which give rise to branching patterns are of the order of one cell size and the interesting range of wave numbers will be small as we are not interested in perturbations of wave length smaller than a cell size ($q \leq 2\pi$). This means that this assumption will be valid within the dynamically interesting range.

The equation of the perturbed nutrient field can now be written as
\begin{equation}\label{eq:perc}
c_\delta(x,\xi ) = \hat c(\xi) - \delta B e^{ - q\left( {\xi  + d} \right)} \cos qx. 
\end{equation}
where the linear part $\hat c(\xi)$ is given by
\begin{equation}
\hat c(\xi)=1-k/v_p^2(1-e^{-v_pd}) + \xi/d(1-k/v_p^2(1-e^{-v_pd})-c_n ) 
\end{equation}
and $B=\hat c'(\xi)=$ constant is determined from the boundary condition $c_\delta(x,-d(x))=c_n$. This field satisfies $\nabla^2 c = 0$  and the boundary condition $c_\delta(x,-d(x))=c_n$ (to first order in $\delta$) and is therefore an approximate solution for the perturbed interface.

\begin{figure}[!hbp]
   \centerline{\includegraphics[width=8cm]{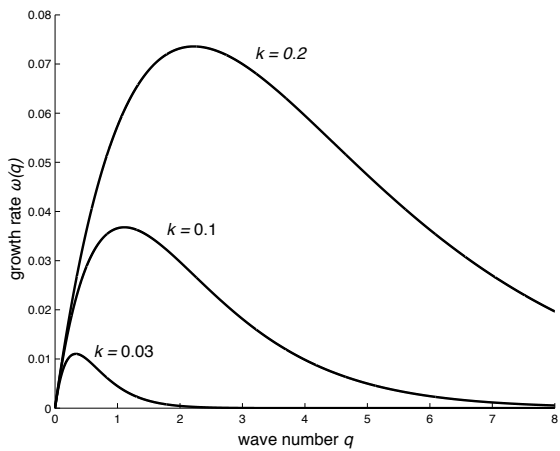}}
\caption{\label{fig:growth}This plot shows the dispersion relation (\ref{eq:disp}) for $k=0.03, 0.1, 0.2$ and $\tau=10$, $c_n=0.1$. It can be seen that both the fastest growing mode and its growth rate depends on the consumption rate $k$.}
\end{figure}

As the nutrient field now depends on $x$ the growth of the interface is as argued above proportional to $\vec n \cdot \nabla c_\delta(x,\xi_0(x))$, where $\vec n = (1+\delta^2 q^2 \sin^2qx)^{-1/2}(\delta q \sin qx, 1)$. But as $\delta \ll1$ the interface velocity in the $x$-direction is negligible and the gradient dependent growth velocity can be approximated by
\[
v(x) = A\left. {\frac{{\partial c_\delta(x,\xi )}}{{\partial \xi }}} \right|_{\xi = \xi_0(x)} =
\]
\begin{equation}\label{eq:graddrowth}
 = A\left( \hat c'(\xi(x)) + \delta B qe^{-q(\delta \cos qx +d)} \cos qx \right) 
\end{equation}
where $A$ is the constant of proportionality. The velocity can also be written as
 \begin{equation}\label{eq:graddrowth2}
v(x) = \frac{{\partial \xi_0 }}{{\partial t}}= \dot \delta(t) \cos qx .
\end{equation}
Taking the derivative in the $x$-direction and equating the two expressions for the velocity gives (only taking into account first order in $\delta$)
\[
\frac{{\partial^2 \xi }}{{\partial t \partial x}} = \left. {\frac{{\partial ^2 c_\delta}}{{\partial \xi \partial x}}} \right|_{\xi  = \xi_0(x)} 
\]
\[
 - \dot \delta q\sin qx =  - AB\delta q^2e^{ - q\left( {\delta \cos qx + d} \right)} \sin qx \approx  -AB \delta q^2e^{ - q d}  \sin qx.
\]
The growth rate $\dot \delta / \delta$ of the perturbation is therefore given by
\[
\omega(q) = \dot{\delta }/\delta  =  ABqe^{ -q d} \sim
\]
\begin{equation}\label{eq:disp}
 \sim kdqe^{-dq}.
\end{equation}
From this we can see that the thickness of the boundary layer $d$ affects the stability of the interface. The wave number which has the highest growth rate is 
\begin{equation}
q_{max}=1/d
\end{equation} 
and when $d$ is large ($k$ is small) only modes with a small $q$ (long wavelength) have a significant growth rate, but for smaller $d$ (larger $k$) the maxima is shifted to larger $q$ (smaller wavelengths) and the growth rates of these wavelengths increase (cf. fig. \ref{fig:growth}).

Qualitatively the dispersion relation (\ref{eq:disp}) can be explained in the following way: A perturbation of the colony interface gives rise to an identical perturbation in the isoconcentration curve $c=c_n$. As the perturbed field is quasi-stationary the perturbations decay exponentially in the direction of the interface (\ref{eq:perc}). The larger the distance $d$ is between the curve $c=c_n$ and the interface, the more the perturbations in the nutrient field decay. Since the interface velocity is proportional to the gradient of the nutrient field this implies that the thicker the boundary layer is the more uniform the interface velocity is, and consequently the interface is less sensitive to perturbations.
 
In the case $d=0$ the dispersion relation reduces to $\omega(q) \sim q$, which is the dispersion relation for Laplacian growth without ultra-violet regularisation \cite{wim}. In this type of growth the interface velocity is proportional to a potential field which is zero on the interface. In the above derivation the field which governs the growth of the interface instead takes on a zero value at a distance $d$ from the interface. 
With the dispersion relation of Laplacian growth in mind we can conclude that by introducing a boundary layer the interface is stabilised for high wave numbers, but that this stabilising effect decreases as the thickness of the boundary layer is reduced. As mentioned before the thickness of the boundary layer $d \sim 1/k$, which means that the stability of the colony growth depends directly on the consumption rate of the individual cells. A low nutrient consumption results in a wide boundary layer that stabilises the growth, while a high consumption gives rise to a thin boundary layer which leads to unstable branched growth.

The reason why all wave numbers have a positive growth rate ($\omega(q) \geq 0$ for all $q$) is because the dynamics do not contain any stabilising mechanism. In the Mullins-Sekerka instability \cite{mullins}, which also describes a diffusion-limited growth, the growth is stabilised by surface tension, which inhibits the growth of perturbations with a large wave number. A similar type of effect can be observed in a reaction-diffusion model describing bacterial growth \cite{wim2}. In the reaction-diffusion model a protrusion into the nutrient side of the interface results in enhanced local growth, but the bacterial diffusion flow is is reduced at the protrusion. It is the relative strength between these two effects that determines the stability of the growth. This type of stabilisation does not occur in our system because the cells are immobile and the growth does not depend on the local curvature of the interface. Consequently there are no perturbations that have negative growth rate.

Another system which exhibits a Mullins-Sekerka like instability is the Fisher equation with a cut-off in the reaction term for low concentrations of bacteria \cite{levine}. This is motivated by the fact that bacteria are discrete entities, which means that at some small concentration the continuum formulation breaks down. Because we consider single cells rather than concentrations, the cut-off effect is already incorporated in our model. On the other hand we also have a cut-off in the growth due to low nutrient concentrations, i.e. no cells divide in regions where $c \leq c_n$. Although this cut-off is due to cells becoming inactive rather than finite particle numbers it might have a similar effect on the stability of the continuous model and is a question certainly worth investigating.

\subsection{Comparison to Simulations}
The above derivation of the dispersion relation (\ref{eq:disp}) contains a number of simplifications and assumptions and it is therefore important to verify the analytic result by comparing it to simulation results from the discrete model. 
This was done by measuring the average branch width in the colony and how it depends on the consumption rate $k$. The consumption rate affects the branching of the colony as it determines the linear stability of the interface. When a branch grows it is constantly subject to perturbations and when it reaches a critical width it becomes linearly unstable and splits, similar to what occurs in splitting of viscous fingers \cite{ris}. As we don't have any detailed information about the dynamics of the tip splitting we will consider a idealised version of the process. We will assume that the branches grow to the critical width $l_c=\lambda_{max}=2\pi/q_{max}$ at which they split and that each splitting gives rise to two branches of equal width. If we assume that no branches are annihilated and that they grow at a constant speed then an estimate of the average branch width in the colony is
\begin{equation}\label{eq:anwidth}
l_{avg} \approx (\lambda_{max}/2+\lambda_{max})/2=3/4\lambda_{max}=3/2\pi d. 
\end{equation}
This is of course a highly idealised picture of the branching process, but contains the essential dynamics as it is clear from figure \ref{fig:col} that the branch width decreases with increasing $k$.

\begin{figure}[!hbp]
   \centerline{\includegraphics[width=4cm]{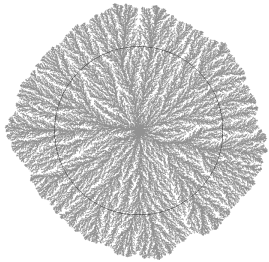}}
\caption{\label{fig:tumour}The circle around which the average branch width was measured.}
\end{figure}

The results from the simulations were analysed in the following way: First the colony was allowed to grow long enough for the morphology to be properly established (approx. 400 time steps), the cell density was then measured on a circle of radius $R=0.75r_{max}$, where $r_{max}$ is the distance to the most distant cell in the colony (cf. fig. \ref{fig:tumour}). The resulting cell density was stored in a vector $n(\theta)$, where $n(\theta)=1$ if the automaton element at distance $R$ and angle $\theta$ from the centre holds a cell (active or inactive) and $n(\theta)=0$ if it is empty and from this vector the average branch width was calculated. In order to make sure that the measurements were not biased by the choice of radius we also measured how the average branch width depended on the radius. The results show that the average branch width depends on the radius for small $R$, but that this bias is negligible for the values of $R$ we consider (data not shown).  

\begin{figure}[!hbp]
   \centerline{\includegraphics[width=8cm]{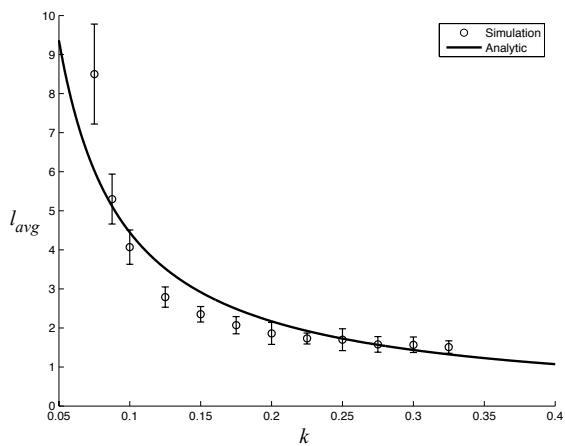}}
\caption{\label{fig:width}The average branch width as a function of the consumption rate $k$. The circles shows the average result from simulations and the solid line represents the analytic result. The error bars show the standard deviation of the simulation results.}
\end{figure}

The average branch width was calculated for several values in the range of $k$ (averaged over 20 simulations for each value of $k$) where branching occurs and the results can be seen in fig. \ref{fig:width}, where it is plotted together with the analytic result (\ref{eq:anwidth}). From this we can see that the average branch width from the simulations agree with the analytic result obtained from the linear stability analysis of the model. The agreement is not perfect but the simulation results exhibit an approximate $1/k$ decay of the average branch width predicted by the stability analysis. One should also bear in mind that our analysis contains a number of simplifications which means that we cannot expect to capture the exact dynamics of the system, but at least our analysis predicts the general behaviour of the system.

\section{Conclusion}
In this paper we have presented a simple hybrid cellular automaton model of cell colony growth, which exhibits interesting growth patterns. We have investigated how the nutrient consumption rate of the cells (or equivalently the nutrient concentration) affects the growth dynamics. The results show that for low consumption rates the colony grows with a Eden-like morphology, while for higher consumption rates the colony breaks up into a branched morphology. By observing that the local growth of the colony is proportional to the gradient of the nutrient field we were able to derive a dispersion relation, which shows that the thickness of the boundary layer in the colony determines the stability of the growth. When the nutrient consumption rate is low the colony exhibits a wide boundary layer, which stabilises the growth, while when the consumption is high the width of the boundary layer is reduced and the growth becomes unstable leading to a branched morphology. When the boundary layer vanishes the derived dispersion relation is reduced to the one describing Laplacian growth without ultra-violet regularisation. 
Analysis of colonies obtained from the discrete model show good agreement between simulations and theory.

Some cells are known to use chemotactic signalling under harsh growth conditions. This has for example been observed in bacterial colonies, which under very low nutrient conditions exhibit densely packed radial branches \cite{walker}. It has been suggested that this occurs because stressed cells in the interior of the colony secrete a signal which repels other cells. This could be included in the model either by introducing a bias towards placing the daughter cell in the neighbouring automaton element that has the lowest (or highest) level of the chemotactic substance or by allowing cells to move down (or up) gradients of the substance. Introducing a chemorepellant secreted by cells exposed to low nutrient concentrations would most likely lead to a more directed growth away from the colony centre, and thus to a more ordered morphology with straighter branches. Another approach could be to introduce a direct chemotactic response to the nutrient, which probably would have a similar effect on the colony morphology. It should be noted that the introduction of chemotaxis would make the dynamics of the model more complex and would render the stability analysis much more difficult.

The instabilities described in this paper should be observable in any system where cell colony growth is limited by some nutrient that diffuses and which leads to inhibition if it falls below some critical threshold. The nutrient field also has to be in a quasi-stationary state, which corresponds to a separation in time scales between the cell division and the dynamics of the nutrient field. Additionally we require that the colony expansion occurs mainly by cell division at the colony boundary and not by movement of individual cells. These conditions apply to growth of avascular tumours, bacterial colonies grown in high agar concentrations, yeast colonies and fungal growth. All of these systems exhibit branched or fractal morphologies under certain growth conditions and these growth patterns may be explained by the dispersion relation presented in this paper.

\section{Acknowledgement} 
We would like to thank Olivier Lejuene and Fordyce Davidson for helpful discussions. This work was funded by the National Cancer Institute, Grant Number: U54 CA113007.

%\bibliography{biblio}

\end{document}